\documentclass{moriond}
\usepackage{hyperref}

\def\be{\begin{equation}}
\def\ee{\end{equation}}
\def\bea{\begin{eqnarray}}
\def\eea{\end{eqnarray}}

\def\GeV{\rm \ GeV}
\def\s{\rm \ s}

\newcommand{\Photo}{\includegraphics[height=35mm]{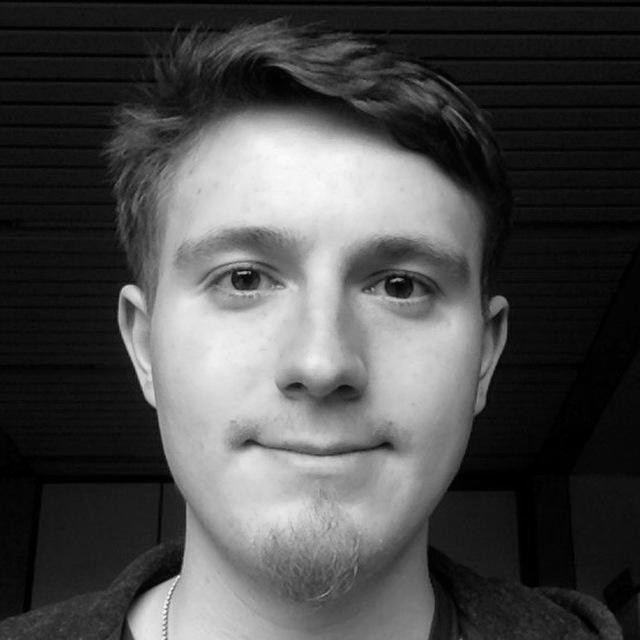}}

\begin{document}
\vspace*{4cm}
\title{CAN DARK MATTER ANNIHILATIONS EXPLAIN THE AMS-02 POSITRON DATA?}

\author{ M. LALETIN }

\address{Space sciences, Technologies and Astrophysics Research (STAR) Institute,\\
Universit\'{e} de Li\`{e}ge, B\^{a}t B5A, Sart Tilman, 4000 Li\`{e}ge, Belgium}

\maketitle\abstracts{
We show that no dark matter model with the conventional isotropic density distribution can provide a satisfactory explanation of the cosmic positron excess, while being consistent with Fermi-LAT data on diffuse gamma-ray background.}

\section{Introduction}
\nocite{Belotsky:2016tja}
In 2009 the PAMELA collaboration had discovered an increase of the positron fraction in cosmic rays at energies above $\sim 10 \GeV$ \cite{Adriani:2008zr}, which was later confirmed by Fermi-LAT \cite{FermiLAT:2011ab} (though with a different normalization factor)  and AMS \cite{Aguilar:2013qda}. This phenomenon known as cosmic positron excess or anomalous abundance of positrons has gained a lot of interest in the theoretical physics society, but despite all the efforts it still remains not fully understandable. A substantial area of all existing explanations is covered by annihilating or decaying dark matter (DM) scenarios (from now on we will refer to this class of models as ``active DM''). In contrast to other explanations the possibility that those high-energy positrons are produced via DM interactions implies that we have finally found the signatures of Beyond Standard Model physics and thus is a tasty morsel for particle physicists. Indeed, in his talk \cite{TingTalkCERN2016} S. Ting, the leader of the AMS-02 collaboration, which has provided so far the most accurate measurement of positron fraction and extended the data to the earlier unexplored energy region, clearly favors the dark matter origin of the observed positron (and also antiproton) signal, but says essentially nothing about the shortcomings of this explanation, namely the constraints on active DM models. The most severe constraints come from cosmic antiproton measurements \footnote{Even in the presence of the high-energy antiproton excess above secondaries \cite{TingTalkCERN2016} (the existence of which is rather controversial \cite{Giesen:2015ufa}) the constraints imposed on active DM are rather strong.} and gamma-ray observations \cite{Cirelli:2016mnb}. And while some of those constraints depend (to an extent) on the particular model of active DM, the constraint we propose is based on just two assumptions: 1) active DM is distributed isotropically in the Galactic halo; 2) it produces the necessary amount of positrons to fit the AMS-02 positron fraction data. The first assumption is a very common feature of active DM models, though we believe that a couple of (so far) exotic models can avoid this constraint (see below). The basic idea of our approach is that the positrons produced in isotropically distributed sources would generate gamma-ray emission, which exceeds the limit set by Fermi-LAT measurement of isotropic diffuse gamma-ray background (IGRB) \cite{Ackermann:2014usa}. 

\section{The IGRB constraint on the positron excess explanations}

We start our analysis \footnote{See also our papers \cite{Laletin:2016egv} and \cite{Belotsky:2016tja} (section 4).} by obtaining the model-independent DM injection spectrum of positrons needed to fit the data.
To calculate the propagation of positrons in the Galaxy we use the GALPROP code \cite{GalpropCode} with the propagation parameters, which provide the best fit of the AMS-02 data on proton flux and $B/C$ ratio in cosmic rays \cite{Jin:2014ica}. Though in principle the predicted flux of positrons is sensitive to the set of propagation parameters this dependence in the high-energy region (above $10 \GeV$) is very subtle (for example, see Fig. 13 in \cite{Cirelli:2010xx}) and we think that our choice of propagation parameters is well justified. A back-of-the-envelope calculation of the average positron diffusion length (given the parameter values we use) 
\begin{equation}
\lambda^2 \sim \int_{E_{\rm fin}}^{E_{\rm in}} dE \ D\beta^{-1} \left(\frac{E}{\rm 1 \GeV}\right)^{\delta},
\end{equation}
where $D$ stands for diffusion coefficient, $\delta$ is the power spectral index and $\beta \sim 10^{-16} \GeV \s^{-1}$ is the positron energy loss rate, shows that our results also do not depend on the choice of active DM density distribution (as long as it is one of the conventionally used isotropic DM profiles), since the estimated distance from which those high-energy positrons mostly come to us hardly exceeds 1 kpc. 

\begin{figure}[h!]
\centerline{\includegraphics[width=0.9\textwidth]{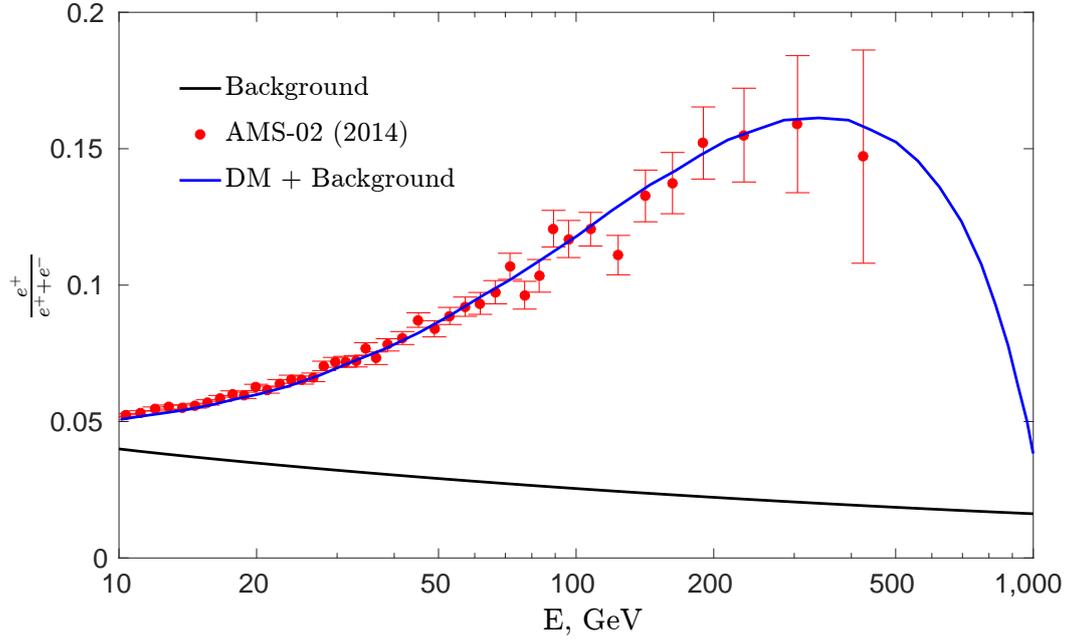}}
\caption[]{Positron fraction resulting from the injection spectrum given by Eq.~\ref{inject} compared to the AMS-02 data~\cite{Accardo:2014lma}. %
} %
\label{positrons}
\end{figure}

We find that a simple power law injection spectrum
\begin{equation}
\frac{dN}{dE} = a \left(\frac{E}{1 \GeV}\right)^{-b}, \ 0 < E < 1 {\rm \ TeV}
\end{equation}
\label{inject}
with $a \approx 3\cdot10^{-8} \GeV^{-1} $ and $b = 1.5$ provides a very good fit of the AMS-02 positron fraction data \footnote{Preliminary AMS-02 data above $500 \GeV$ (see slide 20 in \cite{TingTalkCERN2016}) can be nicely fitted by the considered curve as well.} (see Fig. \ref{positrons}). We set the cut-off scale to 1 TeV (which corresponds to the mass of DM particle in case of annihilations). A larger value of the cut-off energy leads to an even stronger contradiction with the IGRB data since it shifts the predicted gamma-ray spectrum to higher energies where the limit is much lower. 

Using this injection spectrum we now want to calculate the flux of gamma rays emitted by those positrons (see \cite{Laletin:2016egv} for details). Besides Bremsstrahlung and inverse Compton scattering emission appearing as a result of positron interactions with the Galactic gas and electromagnetic medium (CMB, infrared radiation and starlight) we also take into account the final state radiation (FSR) from positron production processes. Thus our estimations set the lower limit on the amount of gamma rays produced in DM models aimed at the explanation of positron abundance. 
Note that in general one would expect a larger contribution to gamma-ray spectrum since building a prompt-radiation-free model is rather complicated and requires fine-tuning.

\begin{figure}[h!]
	\centerline{\includegraphics[width=1\textwidth]{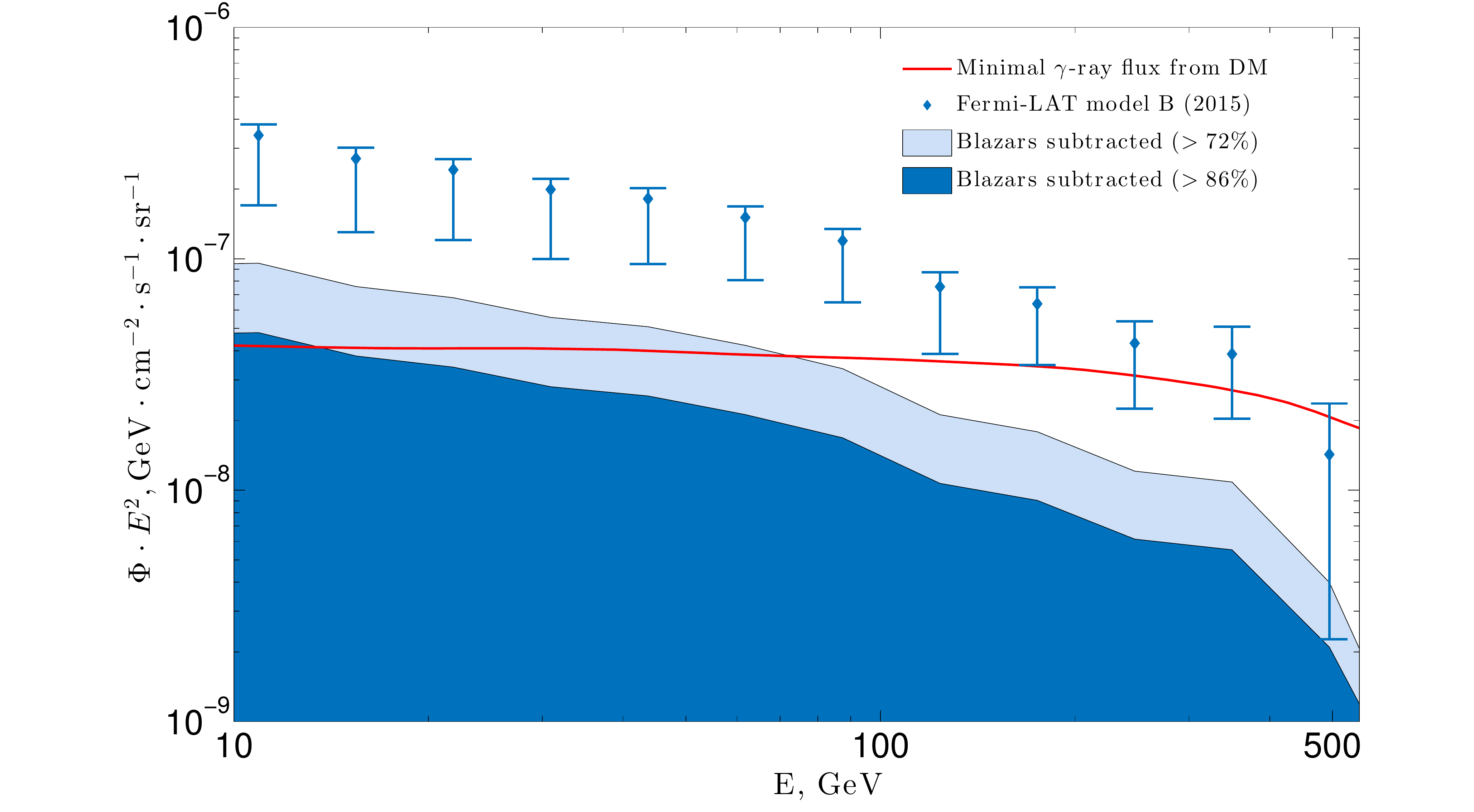}}
	\caption[]{Minimal gamma-ray flux from DM decays or annihilations compared to the Fermi-LAT IGRB data \cite{Ackermann:2014usa} (blue dots). Blue and pale blue areas correspond to the estimated level of IGRB after the subtraction of unresolved AGN contribution (reduced to 28 \% and 14 \% of the IGRB flux, see text).} %
\label{gamma}	
\end{figure}

Finally, we compare the estimated gamma-ray flux to the Fermi-LAT IGRB data \cite{Ackermann:2014usa}. Our goal is not to fit the data points -- they rather set an upper bound on the amount of gamma rays coming from DM decays or annihilations. As one can see in Fig. \ref{gamma} the predicted curve is in good agreement with the Fermi-LAT most conservative foreground model B data. However, the fact that a significant portion of IGRB (around 86\%) can be explained by the contribution of unresolved blazars \cite{TheFermi-LAT:2015ykq} aggravates the situation. Now even the minimal gamma-ray flux from DM is in serious tension with the IGRB limit.   

\section{Discussion}

Our results indicate a strong inconsistency between DM origin of cosmic positron excess and IGRB. In fact, this inconsistency is likely even stronger, since we do not consider neither extragalactic contribution to IGRB from active DM (it depends on the assumptions about density distribution of active DM in the Universe), nor the contribution from ultra high-energy cosmic rays \cite{Kalashev:2016xmy}.

The idea to use IGRB data as a constraint on active DM was studied in a couple of other papers \cite{Liu:2016ngs,Kalashev:2016xmy}, which also seem to disfavor the DM interpretation, though in contrast to us their results are model-dependent. The complementary studies that rule out the DM explanation of positron excess due to the overproduction of gamma radiation include CMB \cite{Slatyer:2015jla} and dwarf galaxy \cite{Ahnen:2016qkx} observations. Note that, however, the former constraint is valid only for annihilating DM and the latter one highly depends on the DM density profile in dwarf galaxies (both of them are also model-dependent). 

To avoid the problem considered in this paper we have proposed a model, in which most of the active DM is concentrated in the dark disk \cite{Belotsky:2016tja}. The model postulates that DM consists of (at least) two fractions, namely ``active'' and ``passive'' DM. The former component being extremely subdominant ($\sim 10^{-3}$ of the cosmological DM density) possesses self-interaction, which lets active DM dissipate energy and form a disk similarly to ordinary matter. This mechanism ensures that annihilations take place only within the disk (which can be up to $\sim 1$ kpc thick) and thus the contribution to isotropic gamma radiation is significantly suppressed. This model has shown no evident contradiction with other gamma-ray observations either \cite{Belotsky:2017wgi}, however a more detailed analysis is required. 

\section*{Acknowledgments}

The author thanks all the co-authors of the original article for their valuable contributions: K.~Belotsky, R.~Budaev and A.~Kirillov. I am also grateful to J.-R.~Cudell, A.~Bhattacharya and M.~Khlopov for comments and discussions. This work is supported by a FRIA grant (F.N.R.S.).

%

\section*{References}


\providecommand{\href}[2]{#2}\begingroup\raggedright\endgroup

\end{document}